# THE MUTUAL TRANSFORMATIONS OF FRACTIONAL –ORDER AND INTEGER-ORDER OPTICAL VORTICES

C.N. Alexeyev, Yu.A. Egorov, A.V. Volyar

*V.I. Vernadsky Crimean Federal University, Vernadsky Prospekt, 4, Simferopol, 295007, Crimea, Russia*

We consider the problem of singular beams in optics as a part of the general questions of interactions, shaping and transformations of vortex states with fractional topological charges in physics, in particular, in hydrodynamic and quantum mechanic. Starting from the representation of the fractional-order vortex states as a superposition of infinite number of integer-order vortices with distinctive energy distributions (the vortex spectra) we showed that the smooth wave front of the fractional vortex beam can either decay into an asymmetric array of integer-order vortices or, vice versa, the array of optical vortices can be gathered together forming a smooth wave front with a helicoid-shaped phase distribution under propagation in free space. It means, on the one hand, that the fractional-order states are structurally unstable. However, on the other hand, the mutual transformations of the fractional vortex states depending on variations of the beam parameters point out that their space-invariant beam behavior can be provided by an appropriate chaise of symmetry and space-variant birefringence of a non-uniform medium.

We revealed that a simple superposition of a finite number of the fractional-order vortex beams enables us to shape symmetric singular beams with the integer-order vortices of the desired topological charges. As such a beam propagates in free space, the centered high-order vortex preserves the state while the vortex framing changes steadily its structure,

We found that the propagation of the fractional-order vortex beams are unstable in biaxial crystals under the condition of the conical diffraction and in the so-called q-plates although the space-variant birefringence of the medium corresponds to the standard polarization distribution of typical representative of fractional order vortex-states – erf-G beams.

At the same time, we revealed that the discrete circular fiber array with a space-variant birefringence is an appropriate medium for fractional-order vortex beams where eigen supermodes bear the half-integer order vortices. In such a medium the standard vortex beams is, in turn, structurally unstable being a superposition of the eigen supermodes. The detail analysis showed that the supermode stability is ensured by two processes: the propagating and evanescent waves (a fiber coupling). The key part in the process plays the non-adiabatic following effect.

## I. INTRODUCTION

The unexpected prediction of vortices with half of a quantum unit of circulation in superfluid $^3He$ in the 1970s [1] at first provoked bewilderment among physicists because the problem was far from an obvious understanding. The fact is that the quantized circulation is connected with a superfluid flow and can have integer values only. For a long time, that prediction had been considered as a mathematical misunderstanding until J. Jang et. al. [2-4] have experimentally revealed half-quantum vortices in different condensed-matter systems from Bose-Einstein condensates to spin-triplet superconductors. To match the experimental results of the half-integer circulations with a generally accepted conception, the deficient phase $\pi$ (a half-integer order of the circulation) came to be called the "hidden phase" that is not connected with the circulation of the mass current but is induced by the circulation of the spin current in Cooper pairs [4]. In other words, the spin-orbit interaction plays here the key part.

At last, there appeared a new approach for describing such physical processes based on analogy between spinor Bose–Einstein condensates and singular optical systems (see e.g. [5] and references therein).

At the same time, the optical states with fractional orders of the energy circulation are not so a drastic problem for flows of optical fields both in scalar and vector cases as it turns out to be for the superfluid cases (Fermi liquids). Nevertheless, propagation processes of singular beams bearing optical vortices with fractional-order topological charges face also the key questions of a structural stability of optical states under propagation or other negligibly small perturbations.

Past recent years of singular optic's [6] development have been marked by the surge of interest to optical vortices with fractional topological charges [7-18]. The first announcement of the fractional-order vortex instability in principle was reported by Soskin et al [7,8] for the vortices produced by a computer-generated hologram while the vortices propagate in free space. The authors observed experimentally evolution of vortex beams with different half-integer order topological charges. If the vortex beam at the hologram has a nearly C-shaped form, far from the hologram the beam breaks out onto a great number of integer-order vortices.

Later, Berry starting from analogy with the Aharonov-Bohm effect in quantum mechanics and hydrodynamics [9] has theoretically shown the splitting of an optical vortex of the fractional-order into infinite chain of the integer-order vortices [10]. He caught sight of a deep analogy between the quantum and the optical singularities. Besides, Berry denoted that the fractional-order vortex propagation results inevitably in decaying the initial phase structure in free space, i.e. the fractional-order vortex beams are structurally unstable ones under the propagation.

These reports stimulated a new chain of theoretical and experimental investigations [11-15] that confirmed the decay of fractional-order vortices into an infinite number of integer-order vortices. Although most of mathematical models of the fractional vortices is based on the Bessel-Gaussian beam presentation (see e.g. [13,14]), authors of the paper [15] supplemented the analysis with the Laguerre-Gaussian beams. Difference between these approaches lies in different contribution of the individual integer-order vortices in the complex field when the fractional-order beam evolves through the optical medium. Some typical features of such dynamical transformations were considered experimentally in the recent paper [18].

On the other hand, authors of the paper [17] found out a strange behavior of the vortex-beam with a half-order topological charge for the erf-Gaussian ($erf - G$) beams. The smooth wave front of the fractional vortex beam can either decay into an asymmetric array of integer-order vortices or, vice versa, the array of optical vortices can be gathered together forming a smooth wave front with a helicoid-shaped phase distribution.

Authors of the paper [21] remarked also unusual behavior of the orbital angular momentum $l_z$ (OAM). At first glance it seems that the fractional-order vortex topological charge is an indicator of the OAM of singular beams at least records nearest values to its physical quantity. In some first papers [14,27] authors obtained a nearly linear dependence between $l_z$ and a topological charge $p$ on the base of assessed theoretical results. However, the computer simulation of the process and physical analysis [21] revealed a complex behavior of the function $l_z(p)$. Small values of the charge $p < 10$ correspond to a nearly linear dependence $l_z \sim p$ with a small amplitude of oscillations. The growth of the value $p > 10$ results in increasing the amplitude of oscillations between the values $l_z = \text{int} eger(p)$ and $p = 0$. The presented results are evidence of a complex interference coupling between a great number of the integer-order vortices in a fractional-order vortex beam.

One more unexpected property of the fractional-order vortex beams revealed the authors of the paper [16]. They tried to answer experimentally the question: can the fractional-order vortex-beams control the states of the integer-order ones? They achieved a success using two beams: the pump and probe ones.

The pump beam is of a topological dipole field consisting of two ½- order vortices with opposite signs of their charges. The pump beam lays a course in a nonlinear medium for the probe singular beam of a smaller intensity. Changing parameters of the dipole they can steer the state of the probe beam. In fact, the fractional-order topological dipole is not destroyed inside the nonlinear medium forming the waveguide channel for the probe beam.

The example of structural stability of fractional-order vortex beams is a discrete fiber array with supermodes bearing half integer-order vortices [19].

Thus, the fractional-order vortex beams possess of a fertile potential of useful properties that call for a real implementation. The priority problems of researches, from our point of view, can be outlined as follows: 1) the processes responsible for decaying and recovery of the initial structure of the beam wavefront inside the fractional-order vortex beams; 2) shaping the singular beams with preassigned properties on the base of a finite superposition of fractional-order vortex beams; 3) search of optical media where the fraction-order vortex beams are eigen modes (i.e. structurally stable wave constructions under propagation); 4) development of the measurement technique and research of the vortex spectra of the singular beam scattered by different linear and nonlinear media; 5) study of nonlinear interactions between the singular beams with the object of employing in computer technologies and other applications. We will focus our attention only on the first three interrelated problems of the above items.

The aim of our paper is to consider the key features of the fractional order vortex beams in free space and space-variant anisotropic media. Namely, 1) the properties of the beams with half integer-order vortices; 2) the shaping of the integer-order beams and their stability under propagation as a superposition of fractional-order vortex beams; 3) structural stability of the beams in biaxial crystals under the condition of the conical diffraction; 4) properties of the beams in a circular discrete fiber array with a space variant birefringence.

In Sec. II we will treat the propagation of fractional-order vortices in free space and uniform media. Our special interest is optical beams with half-integer order vortices. In addition, we will focus our attention on the mutual transformations of integer- and fractional-order vortex beams and their structural stability.

Eigen states of the conical diffraction process in the form of the fractional-order vortices are studied in Sec. III.

The consideration in Sec. IV is accompanied by analysis of the physical properties of the eigen supermodes bearing the half integer-order optical vortices. We concentrate attention on the non-adiabatic following effect responsible for shaping the supermodes with fractional-order vortices. Also we consider the excitation of the waves with integer-order vortices of the array supermodes.

## II. FRACTIONAL-ORDER VORTEX BEAMS IN FREE SPACE

As a rule, a beam field in a complex optical system contains a lot of simple composition Laguerre- Gaussian (LG), Hermit-Gaussian (HG), Bessel-Gaussian (BG) etc beams bearing an energy-limited optical flux. Each element of these mathematical constructions is an optical vortex-beam bearing an integer-order topological charge. Some of such beam combinations possess of extraordinary properties. The singular beam behavior depends on the energy distribution among the integer-order vortices and their phase parameters i.e. a spectral density $\rho(p)$ of optical vortices. A striking example is a fractional-order vortex beam. Its properties are defined by both the value of the topological charge $p$ and type of its complex amplitude (BG, LG etc).

In the following sub-sections, we set a task to uncover the basic properties of different types of the fractional-order vortex beams and to build up from them the integer-order vortex beams.

### A. The fractional vortex states

Let us consider, at first, typical vector supermodes in free space or a uniform isotropic medium made up of the Bessel-Gaussian beams. We focus our attention on monochromatic wave beams with the carrier frequency $\omega$ that enables us to exploit the vector Helmholtz equation for the vector potential $\mathbf{A}$ under the condition of Lorentz gauge [17]. The electric $\mathbf{E}$ and magnetic $\mathbf{H}$ fields can be defined as

$$\mathbf{E} = ik\left[\mathbf{A} + \frac{1}{k^2}\nabla(\nabla \cdot \mathbf{A})\right], \mathbf{H} = \nabla \times \mathbf{A}, \quad (1)$$

with the wavenumber $k$.

Our interest is the paraxial approximation where $|\partial_z^2 \mathbf{A}| << |k^2 \mathbf{A}|$ so that the longitudinal components $E_z$ and $H_z$ can be expressed in terms of the transverse $\mathbf{E}_\perp$ and $\mathbf{H}_\perp$ ones

$$E_z \approx \frac{i}{k}\nabla_\perp \cdot \mathbf{E}_\perp, \quad H_z \approx \frac{i}{k}\nabla_\perp \cdot \mathbf{H}_\perp, \nabla_\perp \equiv \mathbf{e}_x \partial_x + \mathbf{e}_y \partial_y. \quad (2)$$

For the beam propagating along the $z$-axis of the complex amplitude $\tilde{\mathbf{A}}$ of the vector potential $\mathbf{A} = \tilde{\mathbf{A}}(x,y,z)e^{ikz}$ obeys the paraxial wave equation

$$\left(\nabla_\perp^2 + 2ik\partial_z\right)\tilde{\mathbf{A}}_\perp = 0. \quad (3)$$

The choice of the vector $\mathbf{A}$ is defined by a type of the wave beam. If we take, for example, the vector $\mathbf{A}$ to be directed along the $x$-axis (a linearly polarized basis) $\mathbf{A} = \mathbf{e}_x \tilde{A}_\perp \exp(ikz)$ then the solution to the vector wave equation is reduced to the scalar Eq. (3) for the function $\Psi(x,y,z) = \tilde{A}_\perp$ with the solution [20]

$$\Psi = NF(X,Y)G(x,y,z), \quad (4)$$

where

$$G(x,y,z) = \exp\left(i\frac{kr^2}{2Z}\right)/Z \qquad (5)$$

stands for the Gaussian envelope, $Z = z - iz_0$, $z_0 = kw_0^2/2$ is the Rayleigh length with the radius of the beam waist $w_0$, $X = x/Z$, $Y = y/Z$, $N = w_0 \exp\left(-\frac{K^2}{2ikZ}\right)$, $r^2 = x^2 + y^2$ and $K$ is the arbitrary beam parameter that can take on both the real and complex values.

At the same time the function $F(X,Y)$ obeys the two dimensional Helmholtz equation

$$\left(\frac{\partial^2}{\partial X^2} + \frac{\partial^2}{\partial Y^2} + K^2\right) F = 0. \qquad (6)$$

In the cylindrical coordinates the solution to the Eq. (6) can be written as

$$F_p(R,\varphi) = \int_0^{2\pi} \exp\left\{i\left[p\varphi' - KR\cos(\varphi' - \varphi)\right]\right\} d\varphi', \qquad (7)$$

Here the parameter $p \in (-\infty, \infty)$ is arbitrary real value, $R^2 = X^2 + Y^2$. The real part of the parameter $K$ is connected with the half angle $\theta$ of the plane wave's cone of the Bessel beam.

Such a representation of the beam charge $p \in (-\infty, \infty)$ enables us to expand any regular complex beam into the series over different fractional-charged optical vortices just as it can be presented in the form of the series over different integer-order charged optical ones.

To obtain the explicit form of the function $F_p$ in (7) let us use the Fourier transformation

$$e^{ip\phi} = \frac{e^{i\pi p}\sin(\pi p)}{\pi} \sum_{m=-\infty}^{\infty} \frac{e^{im\phi}}{p-m}. \qquad (8)$$

The parameter $p$ here can be regarded as a fractional topological charge, that is responsible for the array of the integer-order vortices with topological charges $m = -\infty \ldots -1, 0, 1, \ldots +\infty$ with the spectral vortex density $\rho(p) = (p-m)^{-1}$. When $p = m$ all terms of the series vanish except the term $e^{im\phi}$. In a general case the function $\rho(p)$ can be defined by a preassigned way as, for example, in the paper [15] for Laguerre-Gaussian beams, but for our purposes we restrict its dependence to the simplest case of $\rho(p) = (p-m)^{-1}$. On the other hand the definition of the Bessel function is

$$2\pi i^m J_m(KR) = \int_0^{2\pi} \exp\left\{i[m\phi + K\cos\phi]\right\} d\phi \qquad (9)$$

with $\phi = \varphi' - \varphi$. As a result, we find

$$|p\rangle = \Psi(r,\varphi,z,p) = 2NG(r,z)\sin(\pi p) e^{i\pi p} \sum_{m=-\infty}^{\infty} \frac{i^m e^{im\varphi}}{p-m} J_m(KR), \qquad (10)$$

Thus, the fractional topological charge $p$ can serve as a global parameter of the complex optical beam. The obtained equation implies two possible propagation processes depending on the value of the $K-$ parameter. The real $K-$parameter is associated with the phase front wreathed by a net of integer-charged vortices at the initial $z = 0$ plane. For example, when propagating, the vortices with $p = 1/2$ begin to form a group in such a way that the vortex net vanishes. There appears the smooth wave front looking like the helix with the phase shift $\Delta\Phi = \pi$ and the C-shaped intensity distribution. For the imaginary value of the $K-$ parameter, the process evolves in the opposite direction [17]. Consider such a process in details.

### B. Asymmetric transverse electric TE and transverse magnetic TM mode beams

**1**. The fractional-order vortex beams permit us to construct unusual wave structures with the broken axial symmetry. In contrast to the usual axial symmetric TE and TM modes with a local linear polarization in each point of the beam, the broken symmetry of the TE and TM mode beams with a fractional order $p = \pm 1/2$ vortices in each polarized component contains local elliptic polarizations at different points of the beam cross-section under the conditions $E_z = 0$ for TE and $H_z = 0$ for TM beams along the beam length. The broken symmetry of the vector field dictates the choice of the basis in the form of circularly polarized components.

From the Eq (2) we obtain for the TE mode ($E_z = 0$, $A_z = 0$)

$$\partial_x E_x = -\partial_y E_y \quad \text{or} \quad \partial_x A_x = -\partial_y A_y \qquad (11)$$

and

$$\partial_x H_x = -\partial_y H_y \quad \text{or} \quad \partial_x A_y = -\partial_y A_x \qquad (12)$$

for the TM mode $(H_z = 0, A_z = 0)$.

It is convenient to employ the circularly polarized basis

$$A_+ = A_x - iA_y, \quad A_- = A_x + iA_y \qquad (13)$$

and a beam vortex structure needs new complex coordinates

$$u = x + iy = re^{-i\varphi}, \quad v = x - iy = re^{i\varphi}. \qquad (14)$$

so that

$$\partial_u = \partial_x - i\partial_y = \frac{e^{-i\varphi}}{2}\left(\partial_r - \frac{i}{r}\partial_\varphi\right),$$

$$\partial_v = \partial_x + i\partial_y = \frac{e^{i\varphi}}{2}\left(\partial_r + \frac{i}{r}\partial_\varphi\right). \qquad (15)$$

Then we find for TE modes $A_+ = \partial_u \Psi_p$, $A_- = -\partial_v \Psi_p$ or

$$E_+ = N\left[\partial_u F_p + ik\frac{v}{2Z}F_p\right]G,$$

$$E_- = -N\left[\partial_v F_p + ik\frac{u}{2Z}F_p\right]G, \qquad (16)$$

where the function $F_p$ obeys Eq. (7).

In optical paraxial cases, where $|\partial_{u,v} F_p| \ll k|F_p|$, we can use the approximation

$$E_+ \approx iNk\frac{v}{2Z}F_p G, \quad E_+ \approx -iNk\frac{u}{2Z}F_p G. \qquad (17)$$

Similarly we obtain TM mode beams

$$E_+ \approx iNk\frac{v}{2Z}F_p G, \quad E_+ \approx iNk\frac{u}{2Z}F_p G. \qquad (18)$$

**2**. Half-order $(2n+1)/2-$ vortex-beams occupy a special place among variety of the fractional-charged optical fields because they can be easily and reliably generated at the initial plane by q-plates [23], photonic crystals [24] and arrays of microchip lasers [25]. Special types of singular beams with the fractional topological charges and fractional orbital angular momentum (OAM) in the closed form (e.g, erf-G beams and others) have been recently considered in number of papers [17,18,21,26].

In this sub-section we will obtain the general closed form of the half-order vortex beams.

As a basic point we take the Eq. (7) and rewrite it in the form of

$$F_p(R,\varphi) = Ke^{i\frac{2n+1}{2}\varphi} \int_{-\varphi/2}^{\pi-\varphi/2} e^{i(2n+1)\phi} e^{-iKR\cos 2\phi} d\phi. \qquad (19)$$

Remember that

$$\cos(2n+1)\phi\, d\phi = \sum_{j=0}^{[n+1/2]}(-1)^j C_n^{2j}\sin^{2j}\phi\cos^{n-2j}\phi =$$

$$= \sum_{j=0}^{[n+1/2]}\sum_{m=0}^{n-j}(-1)^{n-j}C_{2n+1}^{2j}C_{n-j}^m \sin^{2(j+m)}\phi\, d(\sin\phi), \quad (20)$$

$$\sin(2n+1)\phi\, d\phi = \sum_{j=0}^{n}(-1)^j C_{2n+1}^j \sin^{2j+1}\phi\cos^{2(n-j)}\phi =$$

$$= -\sum_{j=0}^{n}\sum_{m=0}^{2j}(-1)^{m+j}C_{2n+1}^j C_{2j}^m \cos^{2(n-j+m)}\phi\, d(\cos\phi),$$

$C_n^m$ – a binomial coefficient.

For example,

$$\cos 3\phi = (1-4\sin^2\phi)d(\sin\phi),$$
$$\sin 3\phi = -(4\cos^2\phi - 1)d(\cos\phi). \quad (21)$$

After substituting (20) into (19) and integrating [22] we obtain

$$F_p = F_n = Ke^{i\frac{2n+1}{2}\varphi}\left\{\sum_{j=0}^{[n+1/2]}\sum_{m=0}^{n-j}(-1)^{n+j}C_{2n+1}^{2j}C_{n-j}^m F_{m,j}^{(s)} + \sum_{j=0}^{n}\sum_{m=0}^{2j}(-1)^{m+j}C_{2n+1}^j C_{2j}^m F_{m,j}^{(c)}\right\}, \quad (22)$$

where

$$F_{j,m}^{(s)} = \frac{\Gamma\!\left(j+m+\frac{1}{2}\right) - \Gamma\!\left(j+m+1/2,\, -2iKR\sin^2\frac{\varphi}{2}\right)}{(-2iKR)^{1/2+j+m}},$$

$$F_{j,m}^{(c)} = -\frac{\Gamma\!\left(j+m+\frac{1}{2}\right) - \Gamma\!\left(j+m+1/2,\, 2iKR\cos^2\frac{\varphi}{2}\right)}{(2iKR)^{1/2+j+m}}, \quad (23)$$

$\Gamma(n,x)$ stands for the incomplete Gamma function.

For example, the fractional beam with $p = 3/2$ is described by the expression

$$\Psi_{3/2} = \frac{NG}{\sigma}K\left\{F_{3/2}^{(s)} + iF_{3/2}^{(c)}\right\}e^{i\frac{3}{2}\varphi},$$

$$F_{3/2}^{(s)} = -\left\{4\sqrt{\Re}\sin\frac{\varphi}{2}e^{-\Re\sin^2\frac{\varphi}{2}} + \sqrt{\pi}(\Re-2)\,erf\!\left(\sqrt{\Re}\sin\frac{\varphi}{2}\right)\right\}/\sqrt{\Re}, \quad (24)$$

$$F_{3/2}^{(c)} = \left\{4\sqrt{-\Re}\cos\frac{\varphi}{2}e^{-\Re\cos^2\frac{\varphi}{2}} - \sqrt{\pi}(\Re+2)\,erf\!\left(\sqrt{-\Re}\cos\frac{\varphi}{2}\right)\right\}/\sqrt{-\Re},$$

$\Re = 2iKR.$

It is useful to mark that the function $\Psi_{3/2}$ in Eq. (24) is a periodic one with the period $2\pi$ despite the factors $\cos\frac{\varphi}{2}$ and $\sin\frac{\varphi}{2}$ in the functions $F_{3/2}^{(c,s)}$. In order to prove it, it is necessary to take into account the factor $e^{i\frac{3}{2}\varphi}$ in the function $\Psi_{3/2}$ and oddness of the function $erf(x)$.

The presented above results are of a new family of asymmetric scalar vortex beams with $p = \pm(2n+1)/2$ that we call Gamma-Gaussian beams ($\Gamma - G$ beams) referring to the complex amplitude $\Psi_p$. The $\Gamma - G$ beams are a natural generalization of the $erf - G$ beams [17] over all set of half integer-order vortex topological charges.

Typical representatives of the $\Gamma - G$ family of the singular beams are shown in Fig. 1. Thus, the field distributions at the beam cross-section depend essentially on the value of the $K-$ parameter. When the $K-$ parameter has a pure real value (see Fig. 1) the intensity distribution has a $C-$ like profile at $z = 0$ with the only half-integer order vortices near the center (see e.g. [17]). However, when propagating the intensity profile is drastically transformed turning into broken Bessel beam at the length $z \gg z_0$ with integer-order vortices scattering over the beam cross-section. For the pure imaginary $K-$ parameter ($|K|$ is constant), the process is reversed.

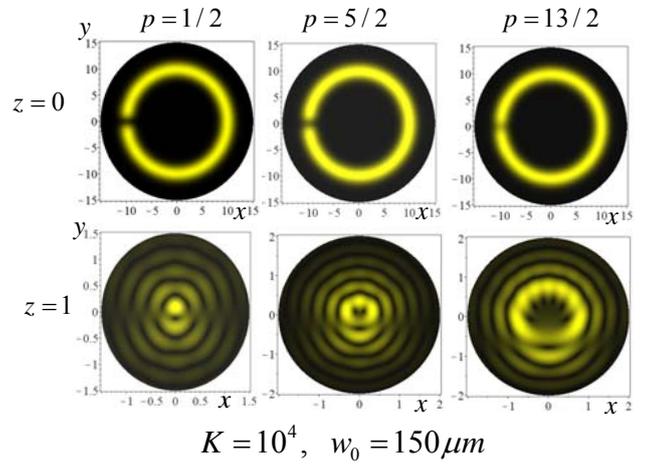

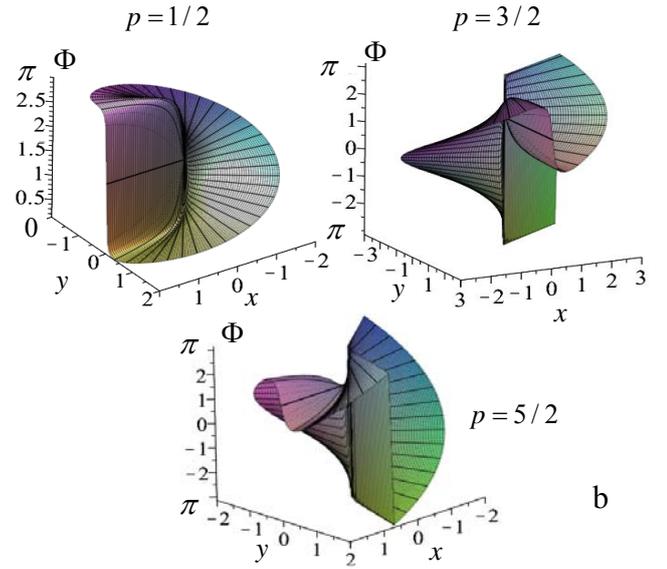

FIG.1. (a) Intensity distributions of the Gamma-Gaussian ($G - \Gamma$) beams with different topological charges $p$ at the initial plane and at the far diffraction zone and (b) their phase distributions as a unified complex field.

The phase distributions shown in Fig. 1(b) illustrate a complex phase structure for different half-order vortex topological charges. A smooth growth of the phase up to $\Phi = \pi/2$ for $p = 1/2$ is replaced by the phase oscillations in the broken second branch of the two-leaved helicoid for the topological charge $p = 3/2$. The phase loss is $\Delta\Phi = \pi/2$. The same phase construction is observed for the topological charge $p = 5/2$ where the third branch of the three-leaved helicoid lacks also the phase $\Delta\Phi = \pi/2$. All phase losses are accompanied by smooth variations. The sign alternation $p \to -p$ changes the direction of the helicoid twist.

All the above equations enable us to build a great number of asymmetric transverse electric *TE* and transverse magnetic TM beams. Some of them are shown in Fig. 2. The fine structure of these fields is reshaped along the beam length, so that the beams are structurally unstable under propagation in free space. In contrast to standard TE and TM modes the asymmetric paraxial beam fields in Fig. 2 are elliptically polarized at each point of the beam cross-section with distinctive orientations of the ellipse axes. Near the optical axis, the field tends to form two polarization singularities of a kind (the star or lemon [12]). Far from the center, the directions of the linear polarization are wound into Archimedean (for TE) and logarithmic (for TM) spirals.

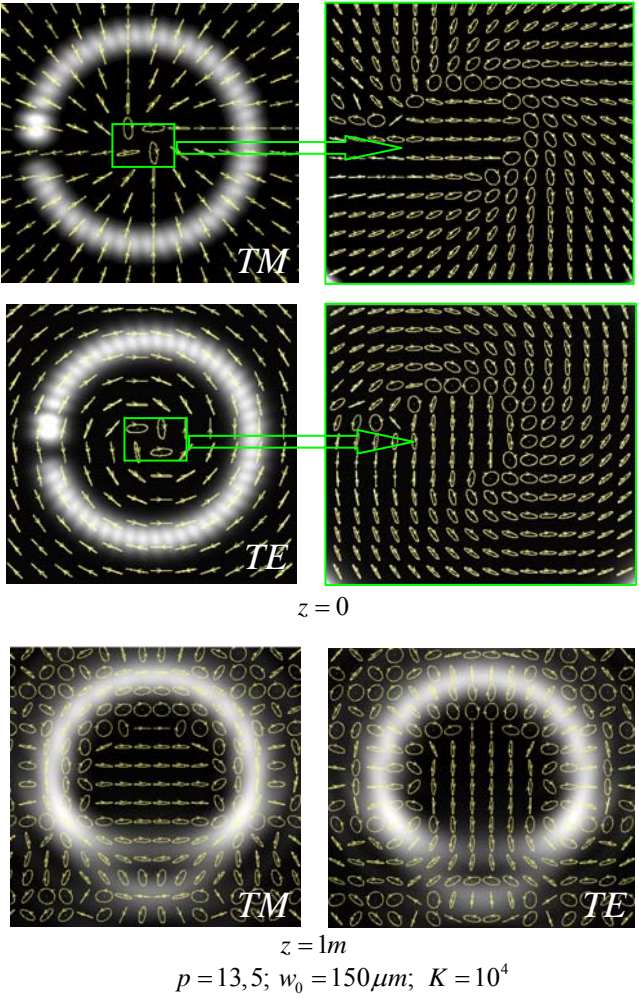

FIG. 2. The field distributions of the $\Gamma - G$ beams for different topological charges at the background of the intensity distributions.

The peculiar feature of the $\Gamma - G$ beams is also their capacity to gather together integer-order vortices into one with the fractional topological charge at far diffraction zone when the $K-$ parameter is a pure real value while a pure imaginary value of the $K-$ parameter induces the reverse process – the fractional vortex decays into an infinite number of integer-order vortices. Such beam behavior reflects the inherent processes in the fractional-order vortex structures in contrast to the representation of the inevitable vortex decaying.

In essence, all types of the considered above fractional order beams are the structurally unstable under propagation.

### C. The shaping of the integer-order vortex beams

Can the fractional-order vortex beams form a stable state of the singular beam with the centered integer-order vortex? This a key question of our consideration here.

At first, we will analyze a dipole structure consisting of two orthogonal states $|p\rangle$ and $|-p\rangle$:

$$|p,-p\rangle = |p\rangle + |-p\rangle = Q \sum_{m=-\infty}^{\infty} \frac{i^m m \, e^{im\varphi}}{p^2 - m^2} J_m(KR), \quad (25)$$

where $Q = 2N G(r,z) \sin(\pi p) e^{i\pi p}$. The state (25) we can regard as the initial topological dipole.

After rotating the initial dipole through an angle $\varphi_q = \frac{\pi}{q}$ (see Fig. 3) so that $\varphi \to \varphi + \frac{\pi}{q}$ we obtain

$$|p,-p,q\rangle = Q \sum_{m=-\infty}^{\infty} \frac{i^m m \, e^{im\varphi}}{p^2 - m^2} J_m(KR) e^{im\frac{\pi}{q}}. \quad (26)$$

Superposition of the Eqs. (25) and (26) gives a new dipole

$$|p,-p,\pm\rangle = |p,-p\rangle + |p,-p,q\rangle =$$
$$= Q \sum_{m=-\infty}^{\infty} \frac{i^m m \, e^{im\varphi}}{p^2 - m^2} J_m(KR)\left(1 \pm e^{im\frac{\pi}{q}}\right). \quad (27)$$

If $q = 1$ the terms with $m = 2m'+1$ for a sign $(+)$ vanish while the residual terms forming the state

$$|p,2,+\rangle = i4Q \sum_{m=0}^{\infty} (-1)^m \frac{(2m)\sin(2m\varphi)}{p^2 - (2m)^2} J_{2m}(KR). \quad (28)$$

In turn, for a sign $(-)$, the terms with $m = 2m'$ vanish and we find the state

$$|p,2,-\rangle = -4Q \sum_{m=0}^{\infty} (-1)^m \frac{[(2m+1)]\sin[(2m+1)\varphi]}{p^2 - [(2m+1)]^2} J_{2m+1}(KR). \quad (29)$$

The first state (28) does not contain any optical vortices but only the set of edge dislocations of the order $p = 2$ as well as the Eq. (29) with $p = 1$.

In order to obtain hight-order beams, e.g. with $p = 4$, we set a phase difference between two dipole states (30) equal to $\Delta\varphi_n = \pi$. As a result one obtains

$$|p,4,\pm\rangle = Q \sum_{m=-\infty}^{\infty} (-1)^m \frac{(2m) e^{i2m\varphi}(1 \pm e^{im\pi})}{p^2 - (2m)^2} J_{2m}(KR) \quad (30)$$

so that the two states are

$$|p,4,+\rangle = i2Q \sum_{m=0}^{\infty} (-1)^m \frac{(4m)\sin(4m\varphi)}{p^2 - (4m)^2} J_{4m}(KR) \quad (31)$$

for the sign $(+)$, and

$$|p,4,+\rangle = i2Q \sum_{m=0}^{\infty} (-1)^m \frac{(4m+2)\sin(4m+2)\varphi}{p^2 - (4m+2)^2} J_{4m}(KR). \quad (32)$$

for the sign $(-)$

By means of such a recurring procedure one finds the general expressions

$$|p,2s\rangle = i2Q \sum_{m=0}^{\infty} (-1)^m \frac{(4sm)\sin(4sm\varphi)}{p^2 - (4sm)^2} J_{4sm}(KR), \; s = 1,2,..., \quad (33)$$

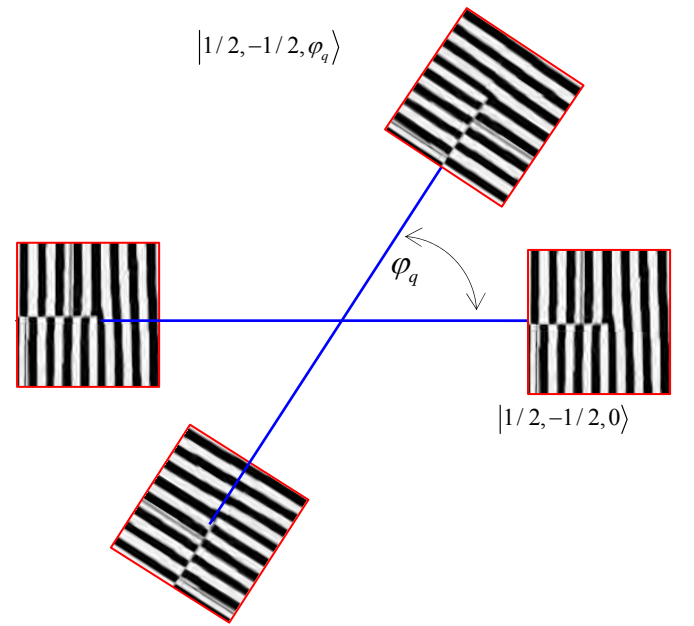

FIG.3. The sketch of the topological dipole and its angular rotation.

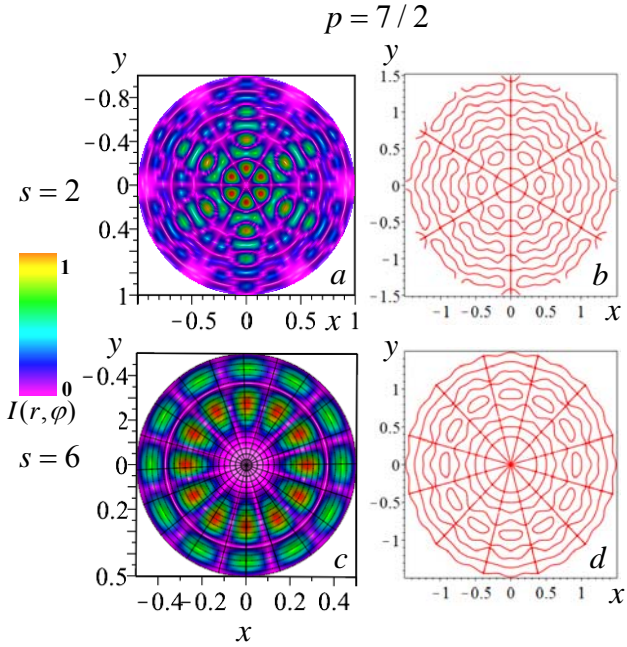

FIG.4. Intensity distributions and L-lines of the axially symmetric beams shaped by the fractional-order vortex beams.

$$|p,2s+1\rangle = -2Q\sum_{m=0}^{\infty}(-1)^m[2s(2m+1)+1]\frac{\sin[2s(2m+1)+1]\varphi}{p^2-[2s(2m+1)+1]^2}J_{2s(2m+1)+1}(KR), \quad (34)$$

$$s = 0,1,2....$$

where $s$ is a number of the recurring transformations, whereas $2s$ and $2s+1$ are topological indices of the wave constructions.

The following step is to rotate the initial dipole through an angle $\varphi_0 = \frac{\pi}{2}$. Such a transformation turns sine in (33) and (34) into cosine at $m=0$ at arbitrary index $s$. As a result we obtain the states with the centered optical vortices of the required integer-order topological charges $l = 2s$ or $l = 2s+1$

$$|p,\pm l\rangle = |p,s\rangle_0 \pm i|p,s\rangle_{\pi/2}. \quad (35)$$

Intensity distributions of the axially symmetric fields shown in Fig. 4(a,c) illustrate the optical constructions built up of the broken fractional vortex-beams on the base of the expressions (33) and (34). Astonishing feature of these structures is that there are no optical vortices in them. Instead we see in Fig. 4(b,d) the curlie-wurlie of the degenerated edge dislocations (see L-lines [48]) webbing tightly around the beam pattern. Three Fig. 4(b) and six Fig. 4(d) nodal lines intersect each other at the axis.

As a result, expressions (35) and (36) are responsible for shaping the optical tracery shown in Fig. 5 consisting of the interchangeable arranged vortex arrays and degenerated edge

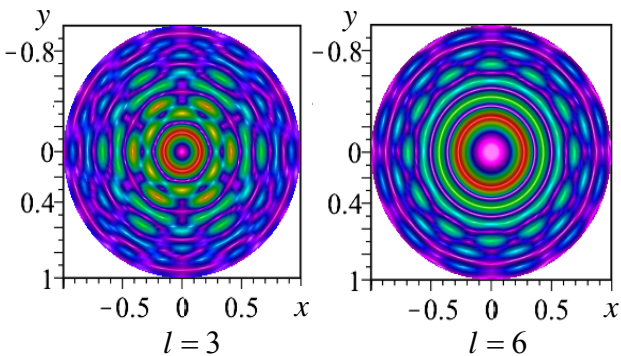

FIG. 5. Intensity beam distributions with centered higher-order vortices and complex vortex framing

dislocations around the centered optical vortex with the topological charges $l = 3$ and $l = 6$.

Thus, a simple rotation of two topological dipoles through discrete angles Fig. 3 enables us to form singular beams with the required, centered integer-order optical vortices. When propagating such a complex beam transforms its framing far from the axis while the central part preserves the singular structure.

At the same time, all the beam states (both with the integer-order and fractional-order vortices) in free space or uniform isotropic media are degenerated, i.e. have the same propagation constants independent on the topological charge values. The choice of the beam representation in one or other basis is determined by the intrinsic symmetry of the optical system. Despite the fractional vortex beam degeneracy in free space, the results presented above enable us to originate new vortex-beam constructions that can open their extraordinary properties in non-uniform anisotropic media as we will see it later.

Also, the problem of the "hidden phase" of the fractional-order vortices [4] in the circularly polarized field components slipped out from our consideration since the fractional-order vortex, as a rule, decays into an infinite number of the integer-order vortices under propagation.

However, the exclusion is the $\Gamma - G$ beams that can either break down the fractional vortex into a set of integer-order vortices or, vice-versa, gather them together into one fractional vortex at far diffraction zone. The control for these counter-directed processes brings into effect the modulation of the beam parameter $K$.

Further we consider two examples of the possible manifestations of the fractional-order vortex beams in the uniform and non-uniform anisotropic media with the distinctive intrinsic symmetry.

## III. THE CONICAL REFRACTION IN BIAXIAL CRYSTALS

### A. General remarks

The fractional-charged beams propagating in uniaxial crystals has been partially regarded in [26] for the vortex beams in states $|\pm 1/2\rangle$ (the so-called erf-G beams). Authors showed conversion between the states $|\pm 1/2\rangle \rightleftarrows |1/2 \pm 2\rangle$ in circular polarized components. It is easily to generalize this rule to arbitrary states $|p\rangle \rightleftarrows |p \pm 2\rangle$. However all the states with different fractional-order vortices are degenerated.

Alternatively, biaxial crystals have one interesting type of the dielectric tensor singularity — Hamilton's diabolical point [28] that defines particular behavior of the vortex beams propagating along one of the crystal optical axes — so called the conical refraction predicted by Hamilton due to a peculiar space-variant birefringence [see Fig. 6]. The internal conical refraction is in spreading a narrow light beam propagating along the crystal optical axis into a hollow cone [29]. The initial circular polarization of the beam splits into a cone of local linear polarizations in such a way that the electric vector **E** rotates though an angle $\pi$ after a full path tracing around the cone axis as shown in Fig. 6. Generalization of Hamilton's approach onto Gaussian beams introduces corrections into a fine structure of the field propagation and distribution [28]. The phenomenon is called the conical diffraction. The conical form of the beam suggested the solutions of the problem in the form of Bessel beams. At the same time, the polarization distribution in Fig. 6 has also much in common with that of

$erf - G$ and $\Gamma - G$ beams in Fig. 4 [17]. Little misalignments of the field patterns far from the optical axis can be referred to a complex structure of the fractional-charged vortex beams.

The unexpected results presented in the papers [29-32] have shown that the uniaxial crystal exhibits a tendency to turn into a biaxial one after twisting it around the optical axis. The space-variant symmetric field of TE or TM eigen modes inherent to a uniaxial crystal [33] at the initial plane $z = 0$ transforms into the asymmetric field distribution similar to that shown in Fig. 6. In contrast to the standard conical diffraction in the typical biaxial crystals, the intensity distribution in the twisted uniaxial crystal has not the pronounced C-shaped form or the circular form with Poggendorff rings [29] but the pattern gets smeared over the cross-section with the singular point at the axis. Nevertheless, the fine structure of the pattern can be controlled by means of either mechanical or electrical devices.

The presented results point out on the fact that eigen mode beams of the conical diffraction and adjoining phenomena are worth searching among the fractional-order vortex-beams.

Thus, the aim of the following sub-section is to study the propagation and conversion of the fractional-order vortex-beams of the Bessel type along one of the optical axes of the biaxial crystal. We will focus our attention to the question: could the input beam field with a space variant polarization identical to that of the crystal birefringence (say, the state $|p\rangle$) propagate without structural perturbation (to be the propagation-invariant wave constructions)? If yes, then we can expect the fractional-order mode beam to be an eigen mode of the medium.

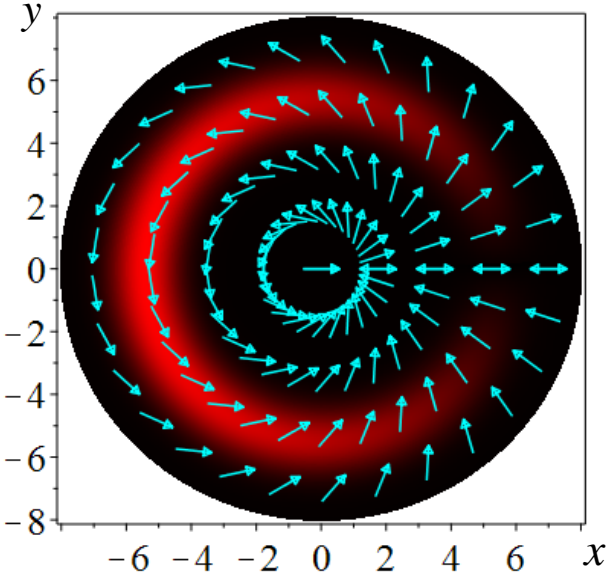

FIG. 6. Space-variant directions of the biaxial crystal birefringence under the conditions of the conical refraction on the background of the beam intensity.

### B. The theoretical treatment

The underlying idea of our treatment leans on the constitutive papers [34-37] where authors consider evolution of the electric field **E** (rather than the electric displacement **D**) of Bessel beams in biaxial crystals under the condition of conical diffraction. The fact is that the wave normal is not directed along the beam propagation in a biaxial crystal so that there appear additional terms in the vector wave equation because of changes in the permittivity tensor. In our case we can use this tensor in the form [37]

$$\hat{\varepsilon} = \begin{pmatrix} \varepsilon_a & 0 & -\varepsilon_{13} \\ 0 & \varepsilon_2 & 0 \\ -\varepsilon_{13} & 0 & \varepsilon_b \end{pmatrix}, \quad (36)$$

where $\varepsilon_a = \varepsilon_1 + \varepsilon_3 - \varepsilon_1\varepsilon_3/\varepsilon_2$, $\varepsilon_b = \varepsilon_1\varepsilon_3/\varepsilon_2$, $\varepsilon_{13} = \sqrt{\varepsilon_1\varepsilon_3(\varepsilon_2-\varepsilon_1)(\varepsilon_3-\varepsilon_2)}/\varepsilon_2$, $n_1^2 = \varepsilon_1$, $n_2^2 = \varepsilon_2$, $n_3^2 = \varepsilon_3$ are the principal refractive indices of the crystal along the axes $x', y', z'$.

The optical axis directed at the angle $\theta$ to the axis $z'$ $\tan\theta = \sqrt{\dfrac{\varepsilon_3(\varepsilon_2-\varepsilon_1)}{\varepsilon_1(\varepsilon_3-\varepsilon_2)}}$. $\varepsilon_1 < \varepsilon_2 < \varepsilon_3$ passes through a diabolic point where slow (s) and fast (f) wavefronts are tangent to each other as shown in Fig. 7.

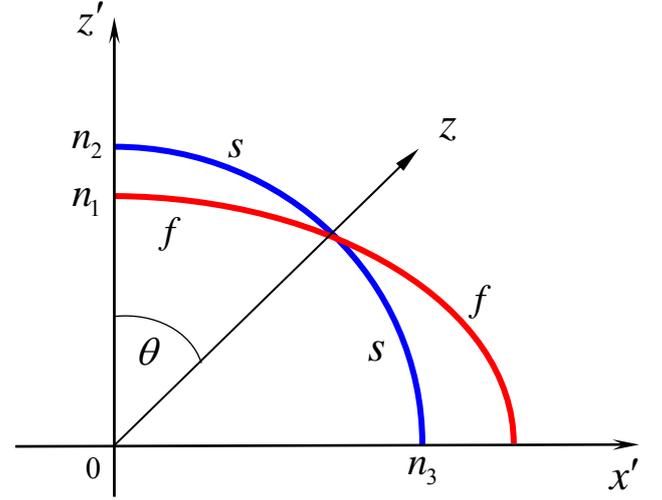

FIG. 7. Sketch of the surface of normals of the slow (S) and fast (f) wavefronts and the optical axis direction.

Author of the papers [35,38] showed that the circularly polarized beam components with the spectral function at the crystal input $A(k_\perp)$ ($k_\perp$ is the transverse wavenumber of the initial beam), are

$$E_+(r,\varphi,z) = \sum_{m'=-\infty}^{\infty} e^{im'\varphi} i^{m'} \int k_\perp A_{m'}(k_\perp) J_{m'}(k_\perp r) \exp\left(-i\frac{k_\perp^2}{2k_b}z\right)\cos(\gamma_0 k_\perp z) dk_\perp e^{i\beta z}, \quad (37)$$

$$E_-(r,\varphi,z) = -\sum_{m'=-\infty}^{\infty} i^{m'} e^{i(m'+1)\varphi} \int k_\perp A_{m'}(k_\perp) J_{m'+1}(k_\perp r) \exp\left(-i\frac{k_\perp^2}{2k_b}z\right)\sin(\gamma_0 k_\perp z) dk_\perp e^{i\beta z}.$$

It means that the right-hand circularly polarized beam bearing a series of the vortex-beams of the m-order and a complex angular spectral distribution $A(k_\perp)$ at the crystal input stimulates an excitation of a series of the vortex-beams of the m+1 – order with the same angular spectrum $A(k_\perp)$ in the left-hand circularly polarized component.

Similarly, it can be shown that the composition of the vortex beams of the m+1 – order with the spectral distribution $A(k_\perp)$ in the left circularly polarized component at the crystal input excites a series of a series of the vortex-beams of the m – order with the same angular spectrum $A(k_\perp)$ in the right-hand circularly polarized component, i.e.

$$E_+(r,\varphi,z) = \sum_{m'=-\infty}^{\infty} e^{im'\varphi} i^{m'} \int k_\perp A_{m'}(k_\perp) J_{m'}(k_\perp r) \exp\left(-i\frac{k_\perp^2}{2k_b}z\right)\sin(\gamma_0 k_\perp z) dk_\perp e^{i\beta z}, \quad (38)$$

$$E_-(r,\varphi,z) = \sum_{m'=-\infty}^{\infty} i^{m'} e^{i(m'+1)\varphi} \int k_\perp A_{m'}(k_\perp) J_{m'+1}(k_\perp r) \exp\left(-i\frac{k_\perp^2}{2k_b}z\right)\cos(\gamma_0 k_\perp z) dk_\perp e^{i\beta z}.$$

The circularly polarized single Bessel beam $E_+^{in} = J_m(k_\perp r)\exp(im\varphi)e^{ik_z z}$ with integer-order topological charge $m$, directed along the crystal optical axis (axis $z$ in Fig. 7), has the conical spectral distribution

$A\left(k_\perp = k_\perp^{(0)}\right) = const$. In order to obtain the beam propagation of such a beam it is sufficient to multiply Eq. (37) by the factor $\delta\left(k_\perp - k_\perp^{(0)}\right)$, and taking into account $m' \to m$ we find

$$\mathbf{E}_1 = \begin{pmatrix} E_+ \\ E_- \end{pmatrix} = \begin{pmatrix} J_m(k_\perp r)\exp(im\varphi)\cos(\gamma_0 k_\perp z) \\ -J_{m+1}(k_\perp r)\exp[i(m+1)\varphi]\sin(\gamma_0 k_\perp z) \end{pmatrix}\exp\left(-i\frac{k_\perp^2}{2k_b}z\right)e^{i\beta z} \quad (39)$$

where $\beta = n_2 z$, $k_b = k n_2/2(1 + \varepsilon_2/\varepsilon_b)$, $\gamma_0 = \varepsilon_{13}/2\varepsilon_b$.

Similar to that we can obtain from Eq. (38) for the initial field in the form $E_-^{in} = -J_{m+1}(k_\perp r)\exp[i(m+1)\varphi]e^{ik_z z}$ the expression

$$\mathbf{E}_2 = \begin{pmatrix} E_+ \\ E_- \end{pmatrix} = \begin{pmatrix} J_m(k_\perp r)\exp(im\varphi)\sin(\gamma_0 k_\perp z) \\ J_{m+1}(k_\perp r)\exp[i(m+1)\varphi]\cos(\gamma_0 k_\perp z) \end{pmatrix}\exp\left(-i\frac{k_\perp^2}{2k_b}z\right)e^{i\beta z}. \quad (40)$$

Combination $\mathbf{E}_1 \pm i\mathbf{E}_2$ of (39) and (40) gives

$$\mathbf{E} = \frac{1}{2}\begin{pmatrix} J_m(k_\perp r)\exp(im\varphi) \\ J_{m+1}(k_\perp r)\exp\left[i(m+1)\varphi - i\frac{\pi}{2}\right] \end{pmatrix} e^{i\left(\frac{k_\perp^2}{2k_b} + \gamma_0 k_\perp + \beta\right)z} \quad (41)$$

and

$$\mathbf{E} = \frac{1}{2}\begin{pmatrix} J_m(k_\perp r)\exp(im\varphi) \\ J_{m+1}(k_\perp r)\exp\left[i(m+1)\varphi + \frac{\pi}{2}\right] \end{pmatrix} e^{i\left(\frac{k_\perp^2}{2k_b} - \gamma_0 k_\perp + \beta\right)z} \quad (42)$$

The Eqs. (41) and (42) show that such fields with space-variant polarization can propagates through the crystal without any structural transformations but with different propagation constants $\beta_\pm = \frac{k_\perp^2}{2k_b} \pm \gamma_0 k_\perp + \beta$.

Difference between the propagation constants $\beta_\pm$ of the mode beams is connected with rotation of the mode fields $\mathbf{E}_1$ and $\mathbf{E}_2$ through an angle $\pi$. Fig. 6 demonstrates the situation when the local directions of the space-variant birefringence $\Delta n(\varphi)$ of the crystal coincides with the local directions of the local field distribution that corresponds the propagation constant $\beta_+$. The field rotation at the angle $\varphi_0 = \pi$ results in changing the sign of the local birefringence $\Delta n(\varphi) \to -\Delta n(\varphi)$ that corresponds to replacement

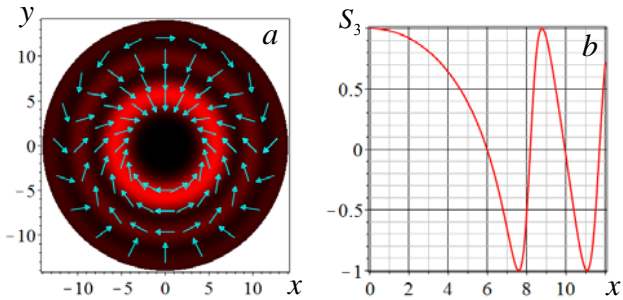

FIG. 8. The polarization state distribution of the mode beam with the index $m = 6$ (a) and (b) the dependence of the ellipticity degree $S_3(r)$ along the beam radius.

$\beta_+ \to \beta_-$.

The polarization state distribution of the s at the mode beam cross-section (6) has a complex form in contrast to the standard structure shown in Fig. 6. A typical space-variant polarization illustrates Fig. 8 for the mode index $m = 6$. Directions of the polarization ellipse axes $\psi$ are depicted on the background of the mode intensity distribution. The ellipticity states specified by the Stokes parameter $S_3$ as a function of radial position in Fig. 8(b) oscillate from the right-hand $S_3 = 1$ to left-hand $S_3 = -1$ states.

However, the ellipticity $S_3$ preserves its value along the azimuth direction $\hat{\boldsymbol{\varphi}}$. Although the path-thracing around the beam axis through an angle $\varphi = \pi$ is accompanied by the ellipse rotation through an angle $\psi = \pi/2$, the full path-tracing results in reinstating both the polarization state and beam phase. Such a space-variant polarization of the eigen mode manifest itself in the ring pattern of intensity distribution while the linear space-variant polarization is in line with $C$-shaped distribution in Fig. 6.

The partial solutions (41) and (42) in the form of eigen modes can be extended to a general solution as a superposition of (41) or (42) with different values of indices $m$. In particular, the mode field $\mathbf{E}_p^{(+)}$ with the propagation constant $\beta_+$ and the fractional topological charge $p$ in the right-hand component can be presented as

$$\mathbf{E}_p^{(+)} = c_p \begin{pmatrix} \sum_{m=-\infty}^{\infty} \frac{i^m J_m(k_\perp r)\exp(im\varphi)}{p-m} \\ -\sum_{m=-\infty}^{\infty} \frac{i^{m+1} J_{m+1}(k_\perp r)\exp[i(m+1)\varphi]}{p-m} \end{pmatrix} e^{i\beta_+ z} =$$
$$= c_p \begin{pmatrix} \sum_{m=-\infty}^{\infty} \frac{i^m J_m(k_\perp r)\exp(im\varphi)}{p-m} \\ -\sum_{m=-\infty}^{\infty} \frac{i^{m+1} J_{m+1}(k_\perp r)\exp[i(m+1)\varphi]}{p+1-(m+1)} \end{pmatrix} e^{i\beta_+ z} = \begin{pmatrix} |p\rangle \\ -c_{p+1}|p+1\rangle \end{pmatrix}\exp(-i\beta_+ z),$$
(43)

where $c_p = \sin \pi p \, e^{ip\pi}$. Similar to that we can write down the mode field $\mathbf{E}_p^{(-)}$ with the propagation constant $\beta_-$:

$$\mathbf{E}_p^{(-)} = \begin{pmatrix} |p\rangle \\ i c_{p+1}|p+1\rangle \end{pmatrix} \exp(-i\beta_- z), \quad (44)$$

with $\beta_\pm = \frac{k_\perp^2}{2k_b} \pm \gamma_{in} k_\perp - \beta$ and we employ the expression (44).

Typical field distributions on the background of the beam intensity is shown in Fig. 9 for the $|0.5\rangle$ and $|7.5\rangle$ fractional states. We calculated the states for the potassium gadolinium tungstate KGd[WO$_4$]$_2$ (KGW) biaxial crystal with refractive indices $n_1 = 2.013, n_2 = 2,045, n_3 = 2.086$, for the wavelength $\lambda = 0.63\,\mu m$, so that the crystal and beam control parameters are $\gamma_{in} \approx 0.0087\,rad$, $\varepsilon_b \approx 4,224$, $k_\perp \approx 1.74 \cdot 10^5\,m^{-1}$. We found the beams in all beam states to have a linear polarization at the cross-section. In contrast to the integer-order vortex charges (see Fig. 8), an angle rotation of the liner polarization $\psi$ of the fractional-order states is multiple to $\pi$ after a full path-tracing around the axis and depends on both the topological charge $p$ and a transverse position $r$. Besides, we found that the greater the value of the parameter $k_\perp$ the greater the number of polarization variations along the radial $\hat{\mathbf{r}}$ directions.

When the initial beam is right-hand circularly polarized, the sought field of the conical diffraction is defined as a combination of the eigen states (43) and (44)

$$\mathbf{E} = \begin{pmatrix} |p\rangle \cos(\gamma_0 k_\perp z) \\ i c_{p+1}|p+1\rangle \sin(\gamma_0 k_\perp z) \end{pmatrix} \exp(-i\bar{\beta}z). \quad (45)$$

where $c_{p+1} = \sin p\pi / \sin(p+1)\pi$.

The beat length in our case equals to $L_B = 2\pi/(\gamma_{in} k_\perp) \approx 4.15\,mm$. It means that the states $|p\rangle$ and $|p+1\rangle$ appear alternately at this length while the eigen mode states (43) and (44) in Fig. 9 emerge at the lengths $L_{e,o} = \pi(2n+1)/(4\gamma_{in} k_\perp), n = 0,1,2,...$.

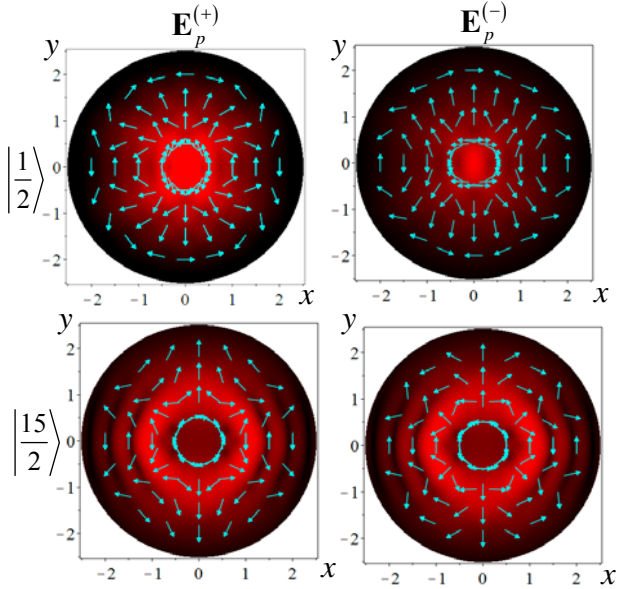

FIG. 9. Field distributions of the fractional-order vortex beam in the potassium gadolinium tungstate KGW biaxial crystal

Thus, the right-hand circularly polarized Bessel beam with the $p-$ fractional-order vortex at the crystal input induces the beam with $p+1-$ fractional-order vortex at the left-hand circular polarization at some crystal length $z$. At the beam length $z = \frac{\pi}{2\gamma k_\perp}(2n+1), \; n = 0, 1, 2, ...$ all energy state is transported from $|p\rangle$ into $|p+1\rangle$ state. However the eigen modes $\mathbf{E}_p^{(\pm)}$ for different charges $p$ have the same propagation constants (i.e. are degenerated). Any superposition of the fractional-order vortex beams obeys the same transformations (45) as the single field states.

Thus, the energy transport of the conical diffraction process in biaxial crystals is carried out from $E_+$ component in $|p\rangle$ state into $E_-$ component in $|p+1\rangle$ state and vice-versa for wide types of the field structure of fractional-order vortex beams. However, difference of the propagation constants between the orthogonal field components is the same for all types of the fractional-order vortex-beams. There is not an appropriate physical mechanism in the biaxial crystals that could make the polarization structure of the beam to follow the singular structure with the fractional-order index of the birefringence directions. As a result the biaxial crystals cannot maintain the single fractional-order vortex beams without their decay into a set of the integer-order vortices.

Let us now peer more attentively into shaping the beam structure in space-variant birefringent media.

## IV. NON-UNIFORM BIREFRINGENT MEDIA

### A. The space-variant unbounded birefringent medium

The brightest representatives of the space-variant media are the so-called q-plates [39]. The q-plate is, in the first version, a slab of a birefringent medium (liquid crystal) with different local directions of the crystal birefringence while the slab has uniform phase retardation. The space-variant birefringence of the q-plates is defined by the topological charge q equal to a rotation of the optical axis in a path circling around the plate center. Obviously, the value of q can be integer or half inter. The q-value can be controlled either by mechanical or by the electrical way [40,41] that implements a polarization modulation at the input beam cross-section. The beam turns into a new wave state due to a superposition of a great number of plane waves with different polarization states. As a result the field distribution has a set of elliptic polarization states differ essentially from the birefringent structure of the q-plate.

There is not an appropriate physical mechanism in the device that could promote imprinting the space-variant birefringence structure into the propagating field. In that respect, the processes of the conical diffraction in the uniform biaxial crystal are not to differ from the effect of the q-plate. In essence, the main mechanism to construct the structured field in the q-plates are the superposition of the uniform propagating waves with the space-variant polarization far from that imprinted by the birefringent distribution of the anisotropic medium while the obtained wave construction maintains the desired fractional angular momentum. It means that the q-plate is solely destined for controlling the orbital angular momentum rather than for creating a stable fractional-order vortex-state.

At first sight it seems that the only physical mechanism of shaping the beams with the space-variant polarization in unbounded media is a superposition of the uniform propagating waves but for one little detail. The Fourier analysis is an appropriate approach only for unbounded media. However, such approach in the paraxial case cannot be applied to the wave beams in restricted media with a boundary surfaces where along with propagating waves exist non-radiative (*evanescent* [52]) waves.

One of such media is photonic crystal fibers that consist of a tightly compressed array of structured optical fibers. Their total birefringence is specified by the structure of a fiber stacking and local properties of single fibers [42-44]. The photonic crystals have two indefeasible advantages: the wave guiding property and the controlled fiber coupling. The simplest model of the photonic crystal is a discrete circular fiber array [45].

In the following section we will try to uncover basic physical processes responsible for the structural stability of vortex constructions with half integer-order topological charges in non-uniform media with a discrete space-variant birefringence and the rules to form from them integer-order vortex beams.

### B. The discrete fiber array: non-adiabatic following and optical quarks

#### 1. Theoretical treatment

We will focus our attention on the discrete system of single mode birefringent fibers inserted into a transparent continuous medium with a uniform refractive index $n_{cl}$ lesser than that of the fiber core $n_{co} < n_{cl}$ [45-47]. Each optical fiber is located at the vertices of a regular *N*-gon as shown in Fig. 10. We will assume that the principal birefringence refractive indices $n_e$ and $n_o$ are such that $n_e \approx n_o \approx n_{co}$, $\delta n = n_{co} - n_{cl} \ll 1$ and $\Delta n = n_e - n_o \ll \delta n$.

The principal point of our consideration is a distinctive distribution of the axes birefringence over the optical fibers: the birefringent directions at $j-$ the fiber makes an angle $\gamma_j^p$ with the $X$ axis of the global frame

$$\gamma_j^p = \frac{2\pi\, p\, j}{N} = 2\varphi_p j, \quad \varphi_p = \frac{\pi}{N} p, \qquad (46)$$

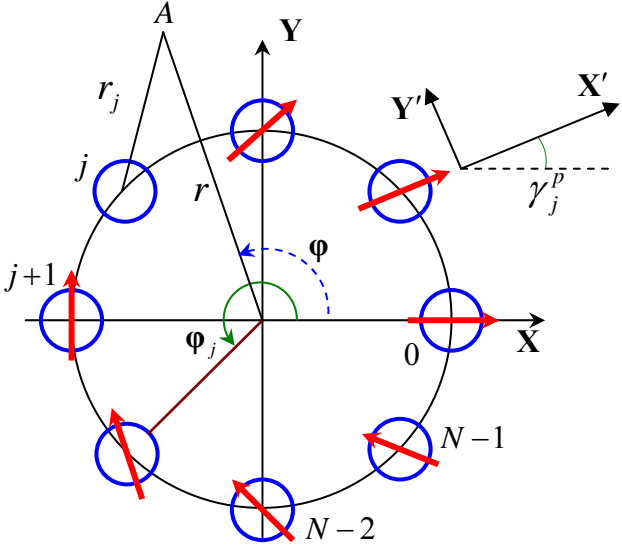

FIG. 10. Sketch of the birefringent fiber positions in the discrete circular fiber array.

where $j = 0,1,2,...N-1$ and $p$ is a number of rotations of the fiber birefringence axis, i.e the index $p$ controls the position of the director of the anisotropic medium. The index $p = (2n_p + 1)/2$, $n_p = 0,1,2,...$ sets the characteristic index of the fiber array. The angle $\varphi_j$ points out the position of the local fiber in the array. Besides our consideration is restricted to the case of even $N$.

The fibers in the array are coupled due to a mutual penetration of the guided fields inside neighboring fibers. The coupling coefficient $a$ (with the dimension $m^{-2}$) is general for all array and depends on the radius of the core, As a result the coupled fiber modes form stable phase-locked field combinations (so-called super-modes) propagating with certain propagation constants. The field structure and the spectrum of their propagation constants are determined by the perturbation matrix [45]:

$$\hat{P} = a\cos 2\varphi_p \begin{pmatrix} 0 & 1 & 0 & ... & 0 & -1 \\ 1 & 0 & 1 & 0 & ... & 0 \\ 0 & 1 & 0 & ... & 0 & ... \\ ... & 0 & ... & ... & 1 & 0 \\ 0 & ... & 0 & 1 & 0 & 1 \\ -1 & 0 & ... & 0 & 1 & 0 \end{pmatrix}, \quad (47)$$

built through averaging over $X',Y'-$ polarized fundamental modes located at individual fibers. The mode spectrum $P_\nu$ is found from the eigenvalue equation

$$\hat{P}\mathbf{K}_\nu = P_\nu \mathbf{K}_\nu. \quad (48)$$

For the components $K_\nu^j$ of the eigenvector $\mathbf{K}_\nu$ one has the following solution

$$K_\nu^j = \frac{\varepsilon^j}{\sqrt{N}}\exp(ij\varphi_{2m+1}), \quad (49)$$

The composite index $\nu$ in (49) consists of two elements $\nu = (\varepsilon, m)$ so that its first element assumes two values: $\varepsilon = \pm 1$, $m = 0,1,..N/2-1$ and the eigenvalue read as

$$P_\nu = \varepsilon a\cos 2\varphi_p \cos\varphi_{2m+1}. \quad (50)$$

The expression for supermodes are built on the basis of the components $K_\nu^j$ and are given by

$$\mathbf{X}_\nu = \sum_{j=0}^{N-1} K_\nu^j \tilde{G}_j \mathbf{i}'_j, \quad \mathbf{Y}_\nu = \sum_{j=0}^{N-1} K_\nu^j \tilde{G}_j \mathbf{j}'_j, \quad (51)$$

where $\mathbf{i}'_j, \mathbf{j}'_j$ are the unit vectors directed along $X', Y'$ axes associated with the $j-$fiber. For the radial function we chose the Gaussian approximation [36]

$$\tilde{G}_j = E\exp\left(-\frac{r_j^2}{2w^2}\right), \quad (52)$$

where $E$ is the field amplitude, $w$ is the waist radius equal $w = \rho_0/\sqrt{2\ln V}$, $\rho_0$ is the radius of the fiber core, $V = k\rho_0\sqrt{2\delta n}$, $k$ is the wavenumber in free space.

The supermodes (51) are formed of fundamental modes of local fibers polarized along $X', Y'$ local axes. The propagation constant $\beta_\nu^{x,y}$ of the $\mathbf{X}_\nu, \mathbf{Y}_\nu$ supermodes is given by [6]

$$\beta_\nu^{x,y} = \bar{\beta} + \frac{P_\nu}{2\bar{\beta}} \pm k\Delta n, \quad (53)$$

where $\bar{\beta}$ stands for the scalar propagation constant of each local fiber. The signs $(\pm)$ denote the upper indices in $\beta^{x,y}$, correspondingly.

Further we will analyze the supermode structure (51) in the circularly polarized basis. Thus, in the general case the mode field of each local mode is elliptically polarized so that the contributions to the $j$-th local fiber make the right-hand circular polarization in the form of the phase factor $\exp\left[i2\pi j(m - n_p)/N\right]$ and the left-hand one in the form of the $\exp\left[i2\pi j(m + n_p + 1)/N\right]$ factor. When we consider the array as a whole, the $j$ index changes from 0 to $N-1$ so that the total increments of the phases over the vertices of the array are $2\pi(m - n_p)$ and $2\pi(m + n_p + 1)$. As we have in detail shown in [46] these increments from opposite circular polarizations set the *integer-order* vortex charge of the discrete fiber array. At first sight it seems that we can conclude that such fiber array *cannot* support the propagation of vortex modes with the fractional-order topological charges. However, we have showed in the Sec. IV(A) that fractional-order vortices can be formed by the superposition of the integer-order vortex modes. It proves also possible to form of the supermodes (49) the simple combinations that explicitly contain the circularly polarized components bearing the fractional-order vortex fields.

The basic point of our consideration lies in choosing the eigen modes bearing the fractional-order vortices. We can reach the desired results through combining the degenerated modes of the fiber array. In fact, the eigenvalues of the matrix $\hat{P}$ in (47) are double-degenerate because $P_{\varepsilon,m} = P_{-\varepsilon, N-2-m}$ [see Eq. (50)] Since $\mathbf{K}^*_{\varepsilon,m} = \mathbf{K}_{-\varepsilon, N/2-m-1}$ it follows that $\mathbf{K}^*_\nu$ belongs to the same eigenvalues as $\mathbf{K}_\nu$. Further we have $(\mathbf{K}^*_\nu \cdot \mathbf{K}_\nu) = 0$, i.e. the vectors are linearly independent. Thus we choose new set of eigenvectors in the form

$$\mathbf{e}_1 = \frac{\mathbf{K}_\nu - \mathbf{K}^*_\nu}{2i}, \quad \mathbf{e}_{-1} = \frac{\mathbf{K}_\nu + \mathbf{K}^*_\nu}{2}. \quad (54)$$

The new set of the eigenvectors can be conventionally divided into two parts with $\varepsilon = 1$ for the $\mathbf{e}_1$ eigenvectors and $\varepsilon = -1$ for $\mathbf{e}_{-1}$ ones. In accordance with Eq. (47) we can obtain the alternative representation of the eigenvector components

$$\Gamma_\nu^j = \frac{1}{\sqrt{N}}\begin{cases}\sin(j\varphi_{2m+1}), & \varepsilon=1\\ \varepsilon^j\cos(j\varphi_{2m+1}), & \varepsilon=-1,\end{cases} \quad (55)$$

while the spectrum of the propagation constants remains defined by Eq. (50) and the eigenvectors are recovered by replacement $K_\nu^j \to \Gamma_\nu^j$ in Eq. (51). On the one hand, the function $\Gamma_\nu^j$ is responsible for a number of zeros in eigen modes of the circular fiber array, but, on the other hand, the function $\Gamma_\nu^j$ lodges the field zeros synchronously with the birefringence directions on the concrete local fibers in the array. We will call the fields with $\varepsilon=1$ and $\varepsilon=-1$ the odd $\mathbf{E}_{o,m}$ and the even $\mathbf{E}_{e,m}$ mode beams,

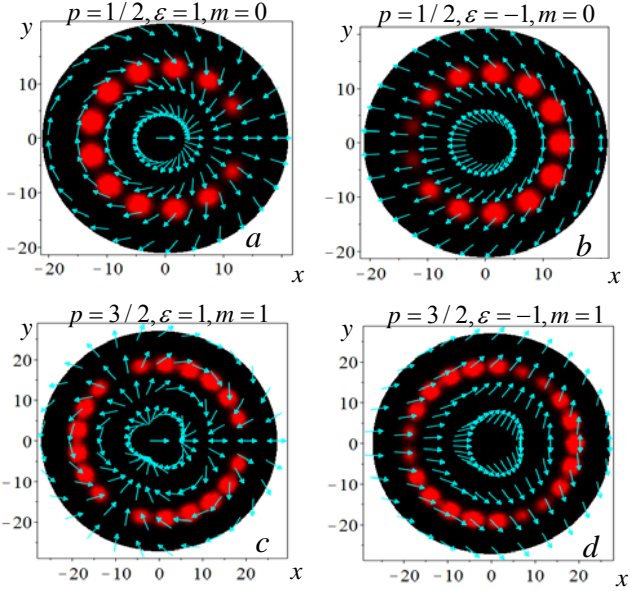

FIG.11. The polarization and polarization distributions for the supermodes with $p=\pm 3/2$.

respectively.

For example, at $\varepsilon=1$ for the amplitudes at the $j-th$ fiber for circular components of the $\tilde{\mathbf{X}}_\nu$ supermode we obtain $\exp(-2i\varphi_p j)\sin(\varphi_{2n+1}j)$ in the right circular polarization and $\exp(2i\varphi_p j)\sin(\varphi_{2n+1}j)$ in the left hand component. For the case $\varepsilon=-1$, the sines should be replaced multiplied by cosines multiple by the factor $\varepsilon^{j-1}$. In this case, the total phase increments in the components over the vertices of the array are $\mp 2\pi p$. In fact, the birefringence symmetry itself of the fiber array inserts the fractional-order topological charges $p$ into supermode fields.

Following [17] we can write for the electric field components $E_\nu^\pm$ of the eigen supermodes:

$$E_\nu^\pm(r,\varphi,z) = G\sqrt{N}\sum_{j=0}^{N-1}\Gamma_\nu^j \exp\left[\frac{rr_0}{w^2}\cos(\varphi-2\varphi_j)\mp 2ip\varphi_i - i\beta_\nu z\right], \quad (56)$$

where $G = E\exp\left[-(r^2+r_0^2)/(2w^2)\right]$, $r_0$ is the array radius.

Typical field patterns on the background of the intensity distributions of the supermodes are shown in Fig. 11. The pattern in Fig. 11(a) has the C-shaped form where the electric field directed along the $\mathbf{X}'$ direction of the birefringence axis in each local fiber (see also Fig. 10) ($|1/2\rangle = E_{o,e}^+$). In the pattern in Fig. 11(b) the intensity distribution is the mirror-reflected intensity in Fig. 11(a). However the electric fields in each local fiber are directed along the $\mathbf{Y}'$ axis ($|-1/2\rangle = E_{o,e}^-$) under the condition that the fiber array index $p$ remains the same (the local birefringent directions do not change). In accordance with the Eq. (53) the propagation constants differ from each other by the value $\Delta\beta = \beta_{1,0} - \beta_{-1,0} = 2k\Delta n$. The patterns in Fig. 11(c) and d have the mirror-reflected positions of the field zeros $(m=1)$ but the fields in each local fiber are directed along the $\mathbf{X}'$ ($|3/2\rangle = E_{1,e}^+$) and $\mathbf{Y}'$ ($|-3/2\rangle = E_{1,e}^+$) axes, correspondingly. The difference between the propagation constants is $\Delta\beta = \beta_{1,1} - \beta_{-1,1} = 2k\Delta n$.

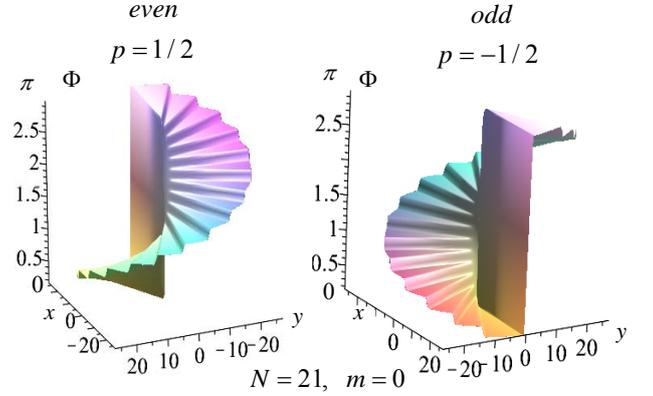

FIG. 12. The ladder-like phase patterns of the supermodes with $p=\pm 1/2$.

Curiously, the point $x=y=0$ around which the full path-tracing is accorded by the rotation of the liner polarization by $\psi=\pi$ is not the singular point in a sense. The fact is that although the field has a space-variant linear polarization over all cross-section, the central point $x=y=0$ cannot be related to any well-known polarization singularities. Typical polarization singularities (star, lemon or monstar) imply the presence of the circular polarization at the center [47].

It is important to note that the local liner polarization in each eigen supermode of the fiber array follows the birefringence axes in the local fibers. Such optical phenomenon has much in common with the phenomenon of the adiabatic following in a twisted birefringent medium (in particular, in liquid crystals) [49]. In contrast to this classical effect, the matching of the field polarization and the fiber birefringence in the discrete fiber array is realized by jumps from one fiber to the other due to the mode coupling from the direction of neighboring fibers. In keeping with the adiabatic following phenomenon in the continuous anisotropic media we call the above effect the *non-adiabatic following* that underlies shaping all eigen supermodes in the discrete circular fiber array.

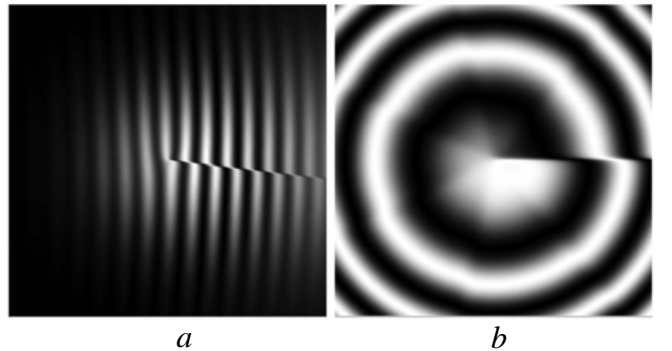

FIG. 13. The interferential patterns of the $E_{o,0}^+$ component with $p=1/2$.

### 2. Is the "hidden phase" hidden indeed?

The following point of our treatment is to study the phase composition in the fractional-order vortex mode components. We plotted the phase patterns using the expressions (56) for the components of the vortex-beams shown in Fig. 12. One observes the ladder-like structure of the phase for the topological charges $p = \pm 1$ for the even and odd field components where the phase jump $\Delta \Phi = \pi$ is present at the center. However, in accordance with our notion in Sec. II(A) the fractional-order vortex beam is an infinite sum of the integer-order vortex beams. It is the set of integer-order vortices that makes up the phase deficiency $\Delta \Phi = \pi$. But we do not observe any traces of the integer-order vortices in Fig. 12.

Perhaps the presented plotting does not feel the hidden vortices? To answer this question and analyze the fine phase structure we studied the interference of the fractional beam with the plane and spherical waves. The interference patterns in Fig. 13 are formed by the superposition of the odd $E_{o,0}^+$ component with the topological charge $p = 1/2$ and the plane (a) and spherical (b) waves. We observe again the ladder-like structure in the phase construction. There is only one broken fork at the end of the cut of the interference fringes [see Fig.13(a)] and the cut of the interference spiral [see Fig.13(a,b)] attesting to the phase jump $\pi$ in the phase structure that resembles the ½-charged vortex imprinted in the field component. There are no any integer-order vortices in the patterns. Besides, such phase distribution preserves its structure when propagating along the fiber array in contrast to the fractional order vortex in free space that is ruined to an infinite number of integer-order vortices [10].

On the other hand, the overall phase increment of $2\pi$, as should have been for a physical meaningful field, is composed of a continuous phase increment of $\pi$ value which we attribute to a ½ - vortex charge and of a phase $\pi -$ jump at the cut of the wavefront that preserves during the mode propagation in the waveguided medium. But such a treatment is only a simple explanation of the interference pattern without a successive physical mechanism that we will consider later on.

At first, let us show that the fractional-order vortex field component has no phase singularities. For this purpose one can study the scalar optical current defined for a scalar field $\Psi$ [17]

$$\mathbf{J} = i\left(\Psi \nabla_\perp \Psi^* - \Psi^* \nabla_\perp \Psi^*\right). \quad (57)$$

Taking into account eq, (56) and (57), one can obtains the expression for the transverse components of the optical current [17]

$$J_x - J_y \propto \sum_{n,m=0}^{N-1} \exp\left[2rr_0 \cos(\varphi - \varphi_{m+n})\cos(\varphi_{m-n})/w^2\right]\sin(2p\varphi_{m-n})\sin\varphi_n \sin\varphi_m \quad (58)$$

The summed expression in (58) is antisymmetric in $m$ and $n$ indices that gives $J_x = J_y$, i.e. the optical current does not contain vorticities. However around the phase singularities the optical current should forms the closed trajectories [48]. Therefore, the circularly polarized component of the fractional-order vortex field (56) does not contain phase singularities.

### 3. Non-adiabatic following and optical quarks

The essential distinction between the continuous and discreet cases is that the smooth field distributions inside the optical fibers are broken by the gaps with other refractive indices and other field nature. In the continuous medium the only propagation wave participates in the transmitting process, at that the beam field of a unified infinite number of integer-order vortices at the initial plane scatters into an infinite number of self-dependent vortices.

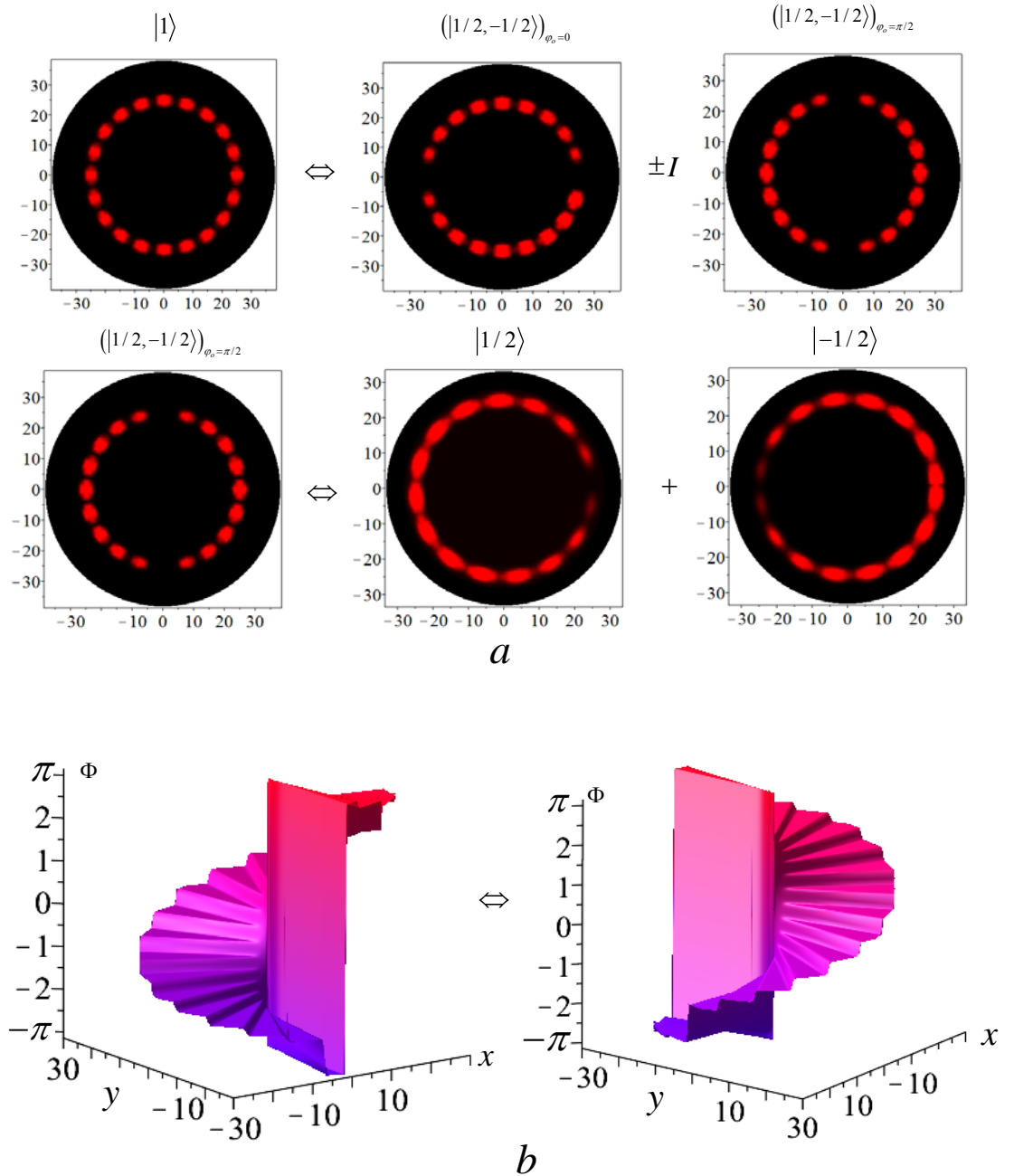

FIG. 14. Diagrams of the conversion of the $|1/2\rangle$ and $|-1/2\rangle$ supermode states.

However, two waves – propagating and evanescent (arising at the refractive index gaps) influence the shaping supermodes of the discreet fiber array. Infinite number of the integer-order vortices in the supermode propagates as a single whole as if the vortices are glued together by the evanescent waves (the fiber coupling) along all length of the fiber array.

A part of the gluing between the fiber fields in the array plays *evanescent waves* [52] (a fiber coupling). It is the evanescent waves that are responsible for *the non-adiabatic following* effect shaping the extraordinary field structure different from that in other propagating fields in continuous media (e.g. in the conical diffraction processes in biaxial crystals and q-plates). The influence of the evanescent waves on the mode shaping falls into place if one imagines that the evanescent waves vanish in a blink. But then the wave guiding effect vanishes too and the supermode turns into an array of divergent light beams with a space-variant polarization so that we return to the q-plate case.

If we traverse around the array center at the radius $r_0$ [see Fig.11(a)] then the inclination angle of the linear polarization in each local fiber changes by jumps from 0 to $\pi$. Such a polarization evolution is mapped on the *Poincare sphere* as a motion along the equator. As a result, each component of the supermode (56) with the topological charge $p=1/2$ acquires the *Pancharatnam–Berry phase* [53] $\phi_{PB}=\pi$ that is of the "hidden phase" considered above.

Let us consider a diagram representation of transmission of the vortex field with integer-order topological charges through a fiber array with half-integer-order index $p$ presented in Fig. 14.

*Case 1.* Let the circularly polarized field with the integer order vortex charge $q=1$ be formed at the input $z=0$ of the array with the index $p=1/2$. Naturally, we must expand the vortex state $|1\rangle$ over the eigen states $|1/2\rangle$ and $|-1/2\rangle$ propagating with their own propagation constants $\beta_q$ and, $\beta_{-q}$ respectively.

The first step to turn the initial vortex field into the topologically neutral one at the arbitrary array length consists in shaping the eigen supermodes $|q\rangle$ and $|-q\rangle$ into two the dipoles – $\left(|1/2,-1/2\rangle\right)_{\varphi_o=0}$ and $\left(|1/2,-1/2\rangle\right)_{\varphi_o=\pi/2}$ rotated through an angle $\varphi_0=\pi/2$ with respect to each other [see the first line in Fig. 14(a)].

The second step is of their superposition with the phase shift $\pi/2$

$$|1\rangle \Rightarrow \begin{pmatrix}\cos\phi \\ i\sin\phi\end{pmatrix}\left[\left(|1/2,-1/2\rangle\right)_{\varphi_o=0} + I\left(|1/2,-1/2\rangle\right)_{\varphi_o=\pi/2}\right], \quad (59)$$

where $\phi = k\Delta n\,z/2$. However the second term in Eq. (59) is nothing but than the sum of the odd and even states

$$\left(|1/2,-1/2\rangle\right)_{\varphi_o=\pi/2} \Rightarrow \left(|1/2,-1/2\rangle\right)^o + \left(|1/2,-1/2\rangle\right)^e. \quad (60)$$

In turn, each topological dipole can be presented as a sum of two modes $|1/2\rangle$ and $|-1/2\rangle$ supermodes [see the second line in Fig. 14(a)]. As far as the supermodes are transmitted through the array with different propagation constants $\Delta\beta = \beta_{1/2} - \beta_{-1/2}$ we will observe the conversion of these states at the beating length $\Lambda = \pi/\Delta\beta$. The conversion of the integer-order states $|1\rangle e^{i\varphi} \Leftrightarrow |-1\rangle e^{-i\varphi}$ in the form of changing the phase starcase $\Phi(x,y)$ is shown in Fig. 14(b).

*Case 2.* Let the transverse electric field ($|TE\rangle$ mode state) be incident on the array input. The input field is of a superposition of two vortex constructions $|1\rangle e^{i\varphi}$, $|-1\rangle e^{-i\varphi}$ that have been regarded in Case 1. The state conversion develops in the frameworks of the above written diagram. We will observe the consecutive alternation of the $|TE\rangle$ and $|TM\rangle$ (transverse magnetic) states at the half-beating length $L_{b/2} = \dfrac{2\pi}{k\Delta n}$ shown in Fig. 15 whereas the conversion of the $|1/2\rangle \leftrightarrow |-1/2\rangle$ states take place at the length $L_{b/4} = \dfrac{\pi}{k\Delta n}$.

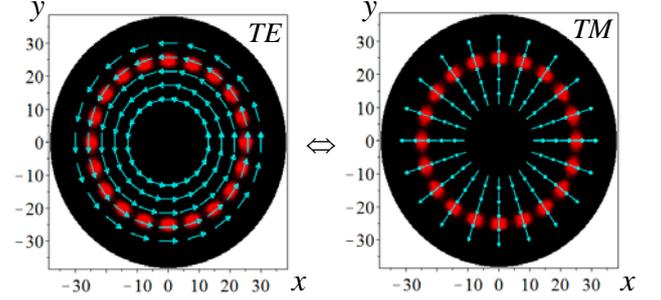

FIG. 15. Diagram of the conversion of the $|TE\rangle$ and $|TM\rangle$ states.

The detailed calculation shows [17,46] that if the projections of the electric field at the array input $z=0$ onto the $X',Y'$ local axes of the $j-th$ fiber are $I_j$ and $L_j$, the amplitudes of the circularly polarized components $A_j^\pm$ at the $\alpha-th$ fiber in the global coordinates $X,Y$ can be written as

$$A_\alpha^\pm = \sum_{v,j} K_v^{j*} K_v^\alpha \left(I_j e^{i\varphi} \mp iL_j e^{-i\phi}\right) e^{i\frac{\Delta\beta^2}{\bar{\beta}}z \mp i\gamma_\alpha^p}, \quad (61)$$

Let a right polarized field with the amplitude

$$A_j^+ \propto \sin n\varphi_j \exp\left(-i2j\varphi_q\right), \quad (62)$$

where the indices $n-$ odd and $q-$ half-integer, be incident onto the fiber array with the index $p$. Then the field amplitude at the $z-$ section of the array is

$$A^\pm \propto G\sqrt{N} \sum_{\alpha=0}^{N-1} A_\varepsilon^\pm \exp\left[\frac{rr_0}{w^2}\cos(\varphi-2\varphi_\alpha)\right] \times$$
$$\times \begin{pmatrix}\cos\phi\, e^{-i\alpha\varphi_{2q}} \\ i\sin\phi\, e^{i\alpha\varphi_{4p-2q}}\end{pmatrix}\sin\left(\alpha\varphi_n - \frac{az}{\bar{\beta}}\cos\varphi_{2p}\sin\varphi_{2p-2q}\sin\varphi_n\right). \quad (63)$$

In fact, the obtained equation is the analog of the expressions (26-28) for the expansion into a series of the field with a fractional-order vortex fields over the integer-order ones in a continuous medium but for the fields in the discrete anisotropic media. When propagating the initial right hand polarized field with the fractional charge $q$ induces the left hand polarized component with the fractional charge $2p-q$ modulated by the factor $\sin\phi$ while the right hand polarization has the factor $\cos\phi$. Otherwise, we observe the energy conversion along the array between the wave fields with the orthogonal circular polarizations but with different fractional topological charges.

Thus, the discrete fiber array with the inherent space-variant anisotropy is a particular medium for translating and preserving light fields with the fractional-order optical vortices. Any other fields bearing the integer-order vortices propagate through the array only as a superposition of the fractional-order eigen supermodes. After the fiber array, the supermode decays into many integer-order vortex beams.

From this point of view, we can regard the fractional vortex states $|p\rangle$ as *optical quarks* (predicted in

[51]) similar to that in the Standard Model of particle physics, in particular, in the Gell-Mann's quark model of the hadrons [50]. The optical quarks in a free state can exist only inside the media with the inherent symmetry of the permittivity tensor. Out of the medium the optical quarks break up into the guided modes of the new optical structure.

## V. CONCLUSION

Different types of symmetry of optical media are the key points that specify properties of the singular beam propagation. It is such starting points that were the base of our consideration. At first, we have considered variety of vector fractional-order vortex beams that can be transmitted through free space or a uniform isotropic medium. Among them the Gamma-Gaussian beams (the $\Gamma - G$ beams, in particular, the erf-G beams) occupy a special place. The fact is that in contrast to the prevalent opinion about decaying the initial fractional-order vortex into a cloud of integer-order vortices, the $\Gamma - G$ beams either break apart of the fractional vortex or, vice-versa, gather together integer-order vortices into one fractional-order vortex at far diffractive zone. However, all types of such vortex beams are unstable under propagation.

On the other hand, we revealed that singular beams with the stable centered integer-order vortices can be formed by four fractional-order vortices. Such constructions remains stable for different values of the topological charges $p$.

We found that the space-variant birefringence with one singular point shown in Fig. 6 is inherent in the fractional-order vortex-beams at the crystals input under the condition of the conical diffraction. Typical scenario of the beam propagation here evolves in such a way that the topological charges of the fractional-order vortices in the circularly polarized components of vector beams differ from each other in one unit. The difference between the propagation constants of the components is independent on the value $p$. It means that the biaxial crystal does not feel distinction between the fractional- and integer-order vortex beams. The same processes we observe also in the so-called *q*-plates, Moreover the polarization states at the beam cross-section are distributed by the complex way far from that of the birefringent directions in the crystal. Naturally the fractional-order vortex beams in the biaxial crystals and q-plates are also unstable one under propagation.

Quite another situation occurs in the discrete circular fiber array. The space variant birefringent axes in the fiber array are exactly recreated at the field cross-section. Such a polarization distribution is well preserved along all fiber array length in the form of a supermode. The shaping of the array eigen supermode is carried out due to the *non-adiabatic following* of the polarization states between the modes of the neighboring fibers when a linear polarization of different parts of the field follows strictly the birefringent axes of the local fibers. Thus, the fractional topological charge of the supermode is specified by the space-variant birefringence of a fiber array. A part of the "glue" of a great number of integer-order vortices into a single fractional-order one plays *the evanescent waves* between the local fibers.

We revealed a remarkable effect of shaping the integer-order vortex in a fiber array. Each integer-order vortex is of a superposition of four fractional order vortices with different propagation constants so that the integer-order mode decays and gathers together again along the array. We came to call them *the optical quarks* owing to resemblance of their behavior with that of quarks in the Standard Model of particle physics. The optical quarks can exist only inside the medium with an appropriate structural symmetry. Outside the medium, the optical quarks are transformed into a cloud of standard integer-order vortices.

## ACKNOWLEDGMENTS

A. Volyar thanks V. Belyi and N. Khilo for the useful discussion on the conical diffraction process in biaxial crystals and E. Abramochkin on the non-factorized paraxial beams.